\documentclass[reprint,aps,pra,amsmath,amssymb,showpacs]{revtex4-1}
\usepackage{graphicx}
\usepackage{dcolumn}
\usepackage{bm}
\usepackage{color}
\usepackage{upgreek}
\usepackage{latexsym}

\begin{document}

\title{Orientation-dependent transition rule in high-order harmonic generation from solids }

\author{Tao-Yuan Du}\email{duty@cug.edu.cn} \affiliation{School of Mathematics and Physics, China University of Geosciences, Wuhan 430074, China}
\author{Si-Jing Ding} \affiliation{School of Mathematics and Physics, China University of Geosciences, Wuhan 430074, China}

\begin{abstract}
We study the role of transition rule in high-order harmonic generation from solids driven by linearly polarized laser fields. The orientation-dependent transition dipoles can regulate the emergence of the multi-plateau structure. In the multi-plateau zone, however, different from the mechanisms of step-by-step excitation reported previously, we find that the emission time and orientation-dependent yields exhibit synchronization and more fine patterns, respectively. And the orientation-dependent yields and its fine patterns can be attributed to the collective contribution of the transition rule and quantum-path interference. Therefore, to better understand the constructive and destructive patterns in the orientation-dependent yields, we propose a scheme of intercycle interference, which can further be utilized as a tool to image the structure of the solids and provide an avenue to optimize the electron dynamics in solids for the production of attosecond pulses in a compact setup.


\end{abstract}

\maketitle

\section{Introduction}

Over several decades, scientists have paid considerable attention on the high-order harmonic generation (HHG) in atomic and molecular systems, which paves a way to generate the table-top light source with energy up to x-ray regimes and leads to the advent of attosecond physics \cite{Krausz,Peng}. In 2011, the realization of efficient HHG from ZnO reported by Ghimire et al. inspired the interest of HHG in solids \cite{Ghimire}. It has quickly expanded into other bulk and layer solid materials, and is realized in liquids recently \cite{Vampa,Ndabashimiye,Tancogne-Dejean,Pronin2,Liu_hanzhe,Yoshikawa,You_NP,Luu,Hohenleutner,Schubert,Lee,Kruchinin,Huttner,Vampa4,Luu_liquid,Heissler}. HHG from solids also provides a novel tool to investigate properties of the crystal materials, such as the reconstruction of the band structure and the Berry curvature \cite{Vampa_recon,Banks_Berry,Luu_Berry,Guan_Berry}. Various physical mechanisms of HHG in solids have been introduced. It can been summarized as only intraband transition  \cite{Ghimire,Pronin2} and the dominated interband transition \cite{Vampa_recon,Banks_Berry,Luu_Berry,Guan_Berry,Hohenleutner,Schubert,Vampa1,Du1,Tancogne-Dejean2,Wu1,Du2,Hawkins,Ikemachi,Du3,ShichengJiang,LiuCD,Hansen,Jin,LiuLu,XiLiu,He_feng,Tamaya,Saito}, or generalized recombination of the electron-hole pair in coordinate and momentum spaces \cite{Vampa1,Du1,Tancogne-Dejean2,Du2,Wu1,Ikemachi,Du3}. Although the mechanisms are intensively disputed, the established picture reaches a good agreement. It can be described as follows: (i) the laser fields pump a small portion of electrons around the top of the valence band into the lowest conduction band; (ii) the electron and hole will oscillate repeatedly in their bands; (iii) population on the higher conduction band starts when the electron is driven to the edge of the Brillouin zone (BZ) within a quarter of laser cycle where the band gap between the first and second conduction band is small. This picture shows that the electron dynamics in the conduction bands is a stepwise process, thus the emissions of transition between the higher conduction and valence bands will be delayed by a quarter of laser cycle relative to the emissions of transition between the lowest conduction and valence bands. As the coherent electron dynamics in our simulations occur at a timescale about a few femtoseconds, faster than the usual timescales of the electron-electron and electron-phonon scattering in solids \cite{Vampa4}, our model do not include these effects \cite{Du1}. The oscillation and polarization of the photoinduced electron-hole pair produce the intra- and interband currents, which lead to the emission of high harmonics.  

When the solid materials are irradiated by the mid-infrared laser pulses, the inter- and intraband currents dominate high harmonic bursts in the plateau and below band-gap zones, respectively \cite{Vampa4,Du1}. In the first step, the field-assisted transition rate which is determined by the transition dipole between the valence and conduction bands greatly influence the subsequent intra- and interband currents and further modulate the high harmonic emissions. In this work, we investigate the impact of orientation-dependent transition dipoles in HHG from two-dimensional (2D) model solid driven by the linearly polarized laser fields. To make an insight into the fine pattern in the orientation-dependent harmonic spectra, we provide a model of intercycle interference. Atomic units are used unless stated otherwise.

\section{THEORETICAL METHODS}

In the single-active-electron approximation, we solve the 2D time-dependent Schr\"odinger equation (TDSE). It can be written as

\begin{equation}\label{E1}
i\frac{\partial}{\partial t}|\Psi(x,y,t)\rangle = (\hat{H}_{0} + \hat{H}_{int}  )|\Psi(x,y,t)\rangle,
\end{equation}
where $\hat{H}_0$ is the time-independent Hamiltonian, and $\hat{H}_{int}$ is the interaction term between the laser and electron of the model solid. They are given by the
$\hat{H}_{0} = \frac{1}{2}\mathbf{\hat{p}}^{2} + V(x,y)$ and $\hat{H}_{int} = \mathbf{A}\cdot\mathbf{\hat{p}}$ in the velocity gauge. $\mathbf{A}$ and $\mathbf{\hat{p}}$ are the vector potential of the linearly polarized laser field and the momentum operator, respectively. $V(x,y)$ is a periodic model potential. In the field-free case, the eigenstate and energy band for the model solids can been obtained by diagonalizing the Hamiltonian $\hat{H}_0$. The form of $V(x,y)$ for one unit cell is
\begin{equation}\label{E2}
V(x,y)=-V_0 \exp\left\lbrace -\left[\alpha_x\dfrac{(x-x_0)^2}{a_{x}^2}+\alpha_y\dfrac{(y-y_0)^2}{a_{y}^2} \right]  \right\rbrace
\end{equation}
where $V_0$ represents the maximum depth of the potential well. Here, we will consider square unit cell with a lattice constant $a$, i.e., $a_x = a_y =a$. $a=4$ a.u., $V_0=3\pi^2/2a^2$, $\alpha_x=\alpha_y=6.5$. ($x_0$, $y_0$) are the coordinates of the center of the potential well. The above model potential does not represent a real crystal unless this system is synthesised in the artificial 2D crystal \cite{Heitmann}, but the conclusion on HHG in this study could be generalized to certain materials with specific symmetry. This 2D system has the point group $C_4$ symmetry  \cite{Jia,Pavelich}. The wavefuction of electron  is mainly localized at each potential well in the valence band, while it becomes more delocalized in the conduction bands.

\begin{figure}[htbp]
\centering
\includegraphics[width=8.5 cm,height=4.25 cm]{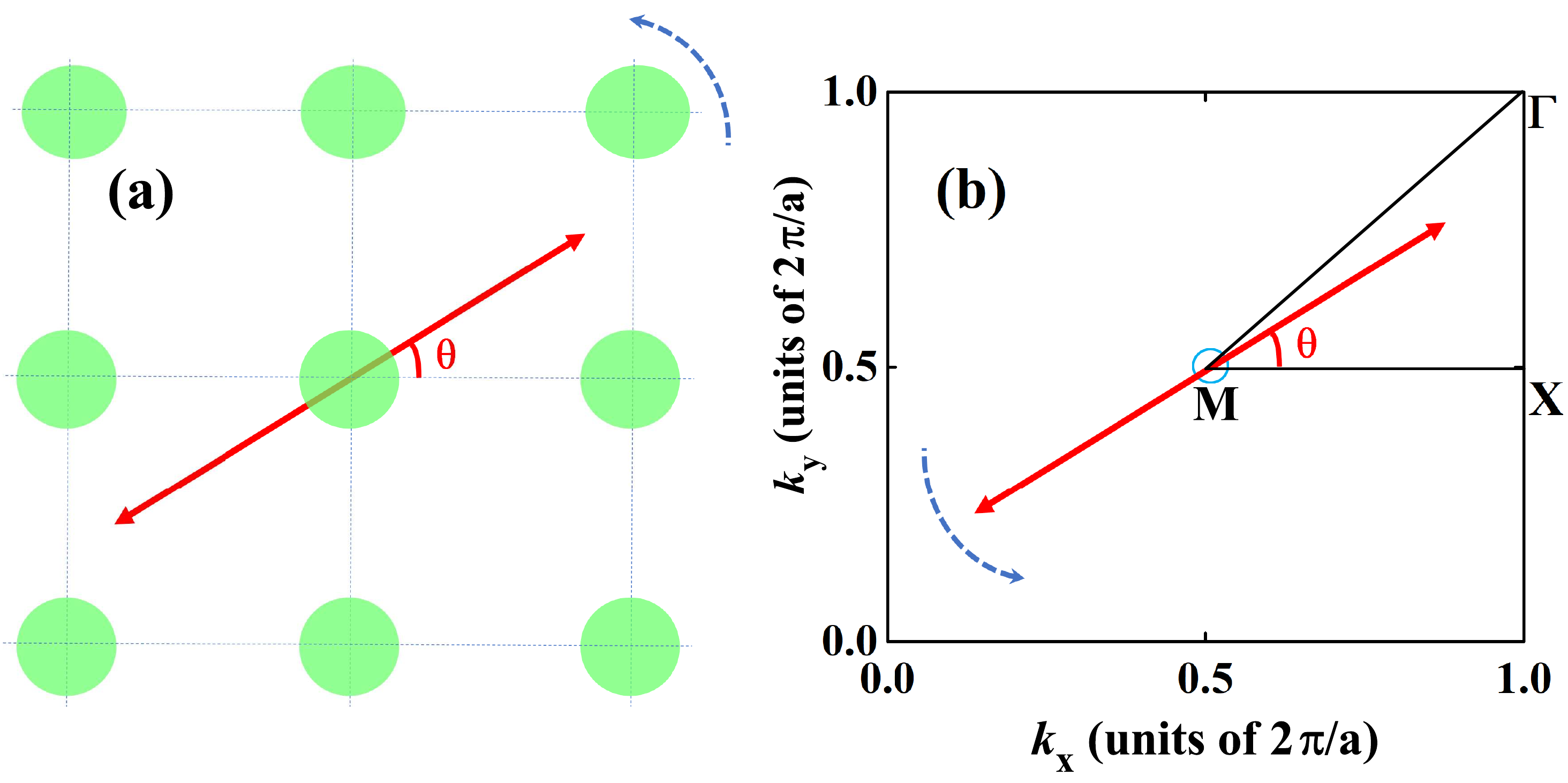}
\caption{ (a) Rotation of 2D solids with respect to the laser polarization direction. The angle $\theta$ shows the crystallographic orientation with respect to the laser polarization direction (red double-headed arrow). (b) The Brillouin zone of 2D solids and high symmetric points M, X, and $\Gamma$ in reciprocal space. M--X (M--$\Gamma$) corresponds to $\theta$ = 0$^{\circ}$ (45$^{\circ}$). The range of the initial states is marked by a circle around M, as shown in (b). }\label{Fig1}
\end{figure}

The time-dependent wavefunction is expanded in a basis of Bloch states, it can been denoted as

\begin{equation}\label{E3}
|\Psi_{\textbf{\emph{k}}}(x,y,t)\rangle = \frac{1}{\sqrt{a_{x} a_{y}}} \sum_{\textbf{n}} C_{\textbf{n}\textbf{\emph{k}}}(t) e^{i[(k_x + \frac{2\pi}{a_x}n_x) x + (k_y + \frac{2\pi}{a_y}n_y) y]},
\end{equation}
where $\textbf{\emph{k}}$($k_x$, $k_y$) are the crystal wave vectors in the selected BZ. The integers $n_{x}$ and $n_{y}$ can be negative, zero or positive. We generate two sequences, $n_{x}$ and $n_{y}$, each extending from $-n_{max}$ to $n_{max}$, and interleave them to produce pairings ($n_x$, $n_y$) with the number of $(2n_{max}+1)^{2}$. We then re-sort the pairings ($n_x$, $n_y$) in an increasing-energy encoding by introducing a new $n_{sort}^{2} = n^{2}_{x} + n^{2}_{y}$ variable and sorting it in ascending order. But for convenience, it is preferable to have a single index $\textbf{n}$ to enumerate the state with the encoding extended from 1 to $(2n_{max}+1)^{2}$. We consider the maximal quantum number ($n_{max} = 3 $) and thus include the energy bands with the number of 49 in our calculations. The transition dipole moments which are described as
\begin{equation}\label{E4}
\begin{split}
  & D_{\textbf{n}'\textbf{n}}(\textbf{\emph{k}}) = i\cdot\frac{\mathbf{\hat{e}}\cdot\mathbf{\hat{p}}_{\textbf{n}'\textbf{n}}}{E_{\textbf{n}'}(\textbf{\emph{k}})-E_{\textbf{n}}(\textbf{\emph{k}})}, \\
  & \mathbf{\hat{p}}_{\textbf{n}'\textbf{n}}(\textbf{\emph{k}}) = \langle u_{\textbf{n}',\textbf{\emph{k}}}(x,y)|\mathbf{\hat{p}}|u_{\textbf{n},\textbf{\emph{k}}}(x,y)\rangle,
\end{split}
\end{equation}
where $|u_{\textbf{n},\textbf{\emph{k}}}(x,y)\rangle$ and $E_{\textbf{n}}(\textbf{\emph{k}})$ are the periodic part of the Bloch state and energy in band index $\textbf{n}$ with crystal momentum \textbf{\emph{k}}, respectively.  $\mathbf{\hat{e}}$ is the unit vector of the linearly polarized laser field, as shown by the red double-headed arrow in Fig. \ref{Fig1}(a). Once the Bloch state is obtained, the values of the transition dipole moments can also be calculated. Here, the matrix elements of transition dipoles have been projected onto $\mathbf{\hat{e}}$ for the different directions.

Considering the transition probability is maximized at the top state of valence band, and the tunneling probabilities rapidly decay with the increase of energy gap. In the HHG spectra presented below, we consider the initial states within the range of $(1\pm5\%)\textbf{\emph{k}}_{0}$ around the high symmetry point M, which can be specified as
\begin{equation}\label{E5}
\textbf{\emph{k}}_{0} = (\frac{\pi}{a}, \frac{\pi}{a}).
\end{equation}
The range of initial states for the different orientations is marked by a circle in Fig. \ref{Fig1}(b). The top state of valence band corresponds to an initial wavefunction which is spatially delocalized. That is similar to the initial condition proposed in Ref. \cite{Wu1}. Please note that the choice for the larger range of initial states in the BZ will give rise to similar conclusions. We assume that the electric field $\textbf{F}=-\frac{\partial \mathbf{A}}{\partial t} $  of the laser pulses with a sine-squared envelope. The wavelength, intensity and duration of the laser pulses are 3.2 $\mu$m, 8.77 TW/cm$^{2}$ and six optical cycles, respectively.

The time-dependent coefficient is obtained using the Crank-Nicholson method \cite{Du3}. The harmonic generation spectrum can be calculated by the Fourier transform of the laser-induced current $J(t) = -\sum_{\textbf{\emph{k}}}[\langle\Psi_{\textbf{\emph{k}}}(t)| \mathbf{\hat{e}}\cdot\mathbf{\hat{p}}  |\Psi_{\textbf{\emph{k}}}(t)\rangle] + A(t)$. Here, $A(t)$ is the vector potential of the laser field which is linearly polarized along unit vector $\mathbf{\hat{e}}$, and the momentum operator $\mathbf{\hat{p}}$ is projected onto $\mathbf{\hat{e}}$. Contributions from different $\textbf{\emph{k}}$ are added coherently.

\begin{figure*}[htbp]
\centering
\includegraphics[width=13 cm,height=9.0 cm]{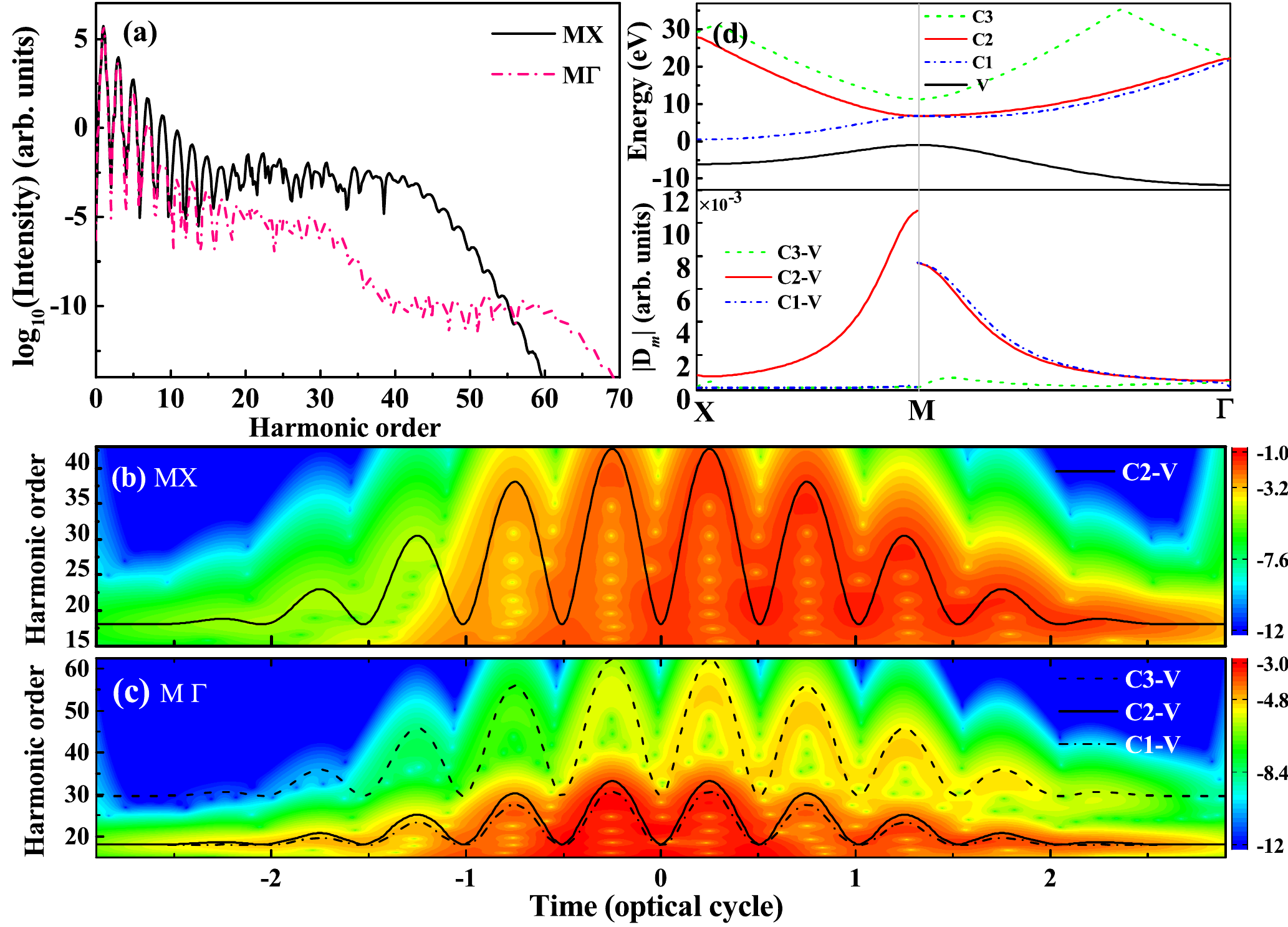}
\caption{ (a) The calculated HHG spectra when the polarization of laser field are along the high symmetry directions M--X (black solid line) and M--$\Gamma$ (red dashed line), respectively. (b) and (c) show the time-frequency analyses of the HHG spectra along M--X and M--$\Gamma$ directions respectively. The lines in (b) and (c) are the quasi-classical predictions \cite{Du1}. Top panel in (d) shows the E--\textbf{\emph{k}} dispersions along the high symmetry direction (X--M--$\Gamma$). The modulus of transition dipole moments between conduction and valence bands are shown in the lower panel of (d).  }
\label{Fig2}
\end{figure*}

\section{RESULTS AND DISCUSSIONS}
Firstly, we make a comparison of the high harmonic spectra when the polarization direction is changed from M--X ($\theta = 0^{\circ}$) to M--$\Gamma$ ($\theta = 45^{\circ}$) by rotating the 2D solids, as presented in Fig. \ref{Fig2}(a). One can observe that the high harmonic spectrum characterizes a rapid decay zone and primary plateau structure along M--X direction. However, a double-plateau structure with a lower yield emerges in the high harmonic spectrum along M--$\Gamma$ direction. To get an insight into the mechanisms of the emergence of double-plateau structure, we will show the time-frequency analysis which can reveal the emission times of high harmonics \cite{Chandre,Du_inhomo}, as presented in the color maps of Figs. \ref{Fig2}(b) and \ref{Fig2}(c). The mechanisms reported previously have revealed that the electrons are promoted by the stepwise transition between the lowest and high-lying conduction bands around the edge of BZ \cite{Du1,Wu1}, thus a quarter cycle of the delay time exists in the emission times for the secondary plateau relative to the primary plateau. However, one can find that the emission times in these two plateaus exhibit a picture of the simultaneous radiations, as shown in Fig. \ref{Fig2}(c).

\begin{figure*}[htbp] \centering
\includegraphics[width=12 cm,height=10 cm]{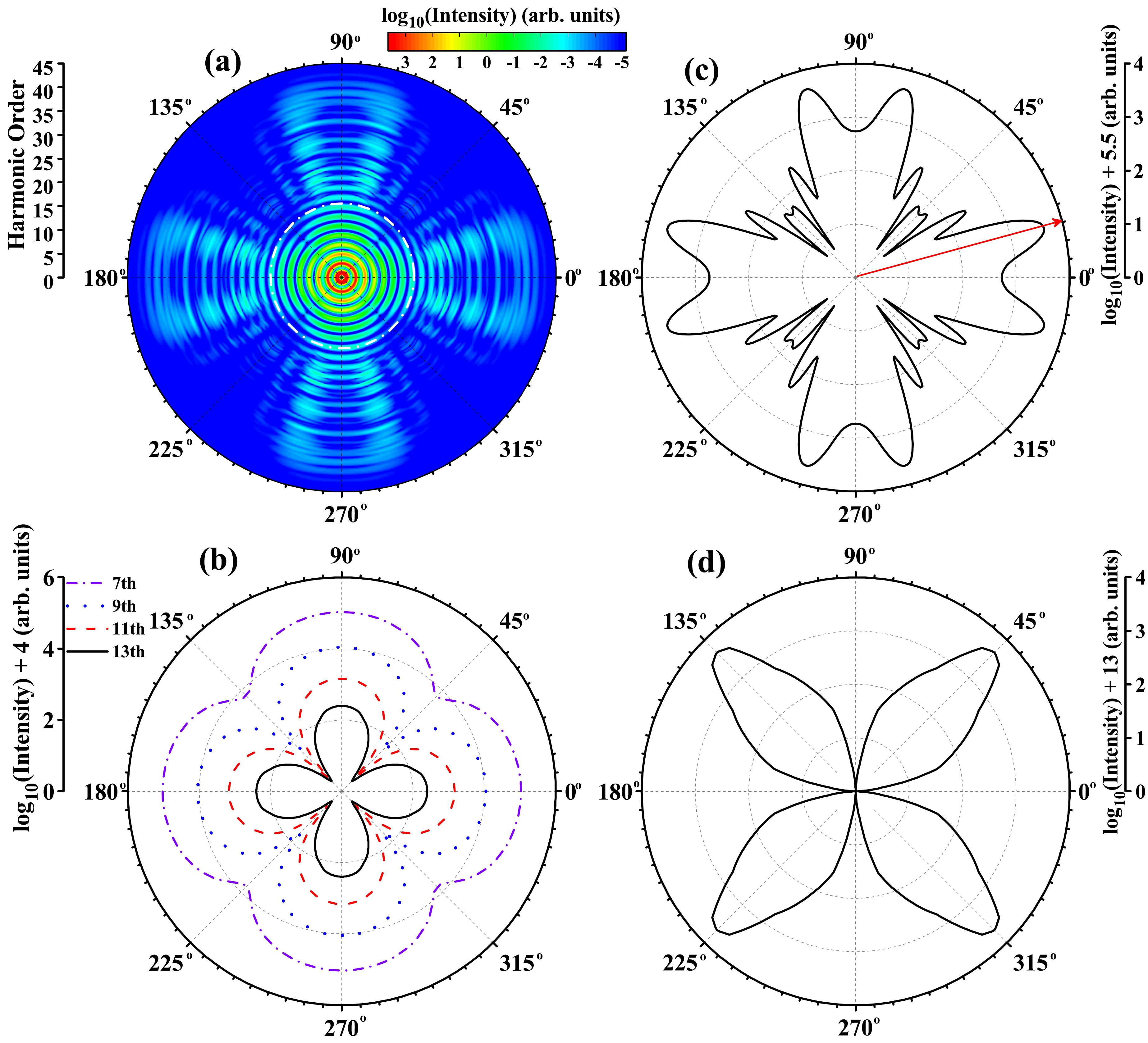}
\caption{ (a) Orientation-dependent high harmonic spectra from 2D solids driven by the linearly polarized laser field. A white dash-dotted circle divides it into below-gap and primary plateau zones. (b) Orientation-dependent yields of the single harmonic in the below-gap zone. In (c) and (d), the orientation-dependent yields of the primary plateau in (a) and the secondary plateau (not shown in (a)) are calculated by integrating the corresponding photon energy ranges respectively.   }\label{Fig3}
\end{figure*}

To make the quasi-classical dynamic analysis, a classification of the energy bands has been provided. We extract the energy-momentum (E--\textbf{\emph{k}}) dispersions along the high symmetry direction (X--M--$\Gamma$) from the 2D energy band structure \cite{Pavelich}, as shown in the top panel of Fig. \ref{Fig2}(d). These E--\textbf{\emph{k}} dispersions are divided into two groups: valence (V) and conduction (C1, C2 and C3) bands. In the lower panel of Fig. \ref{Fig2}(d), we present the absolute values of transition dipoles between the valence and conduction bands based on the calculations in Eq. (\ref{E4}), which uncover the transition rule between the valence and conduction bands along this high symmetry direction. The values of the transition dipoles show that the transitions between all conduction bands and valence band are allowed along M--$\Gamma$, but the only transition between C2 and valence bands is allowed as the laser polarization is rotated along M--X direction. According to the above transition rule and quasi-classical model \cite{Du1}, we can predict the emission times of high harmonics which reach a good agreement with the time-frequency analysis, as shown by the lines in Figs. \ref{Fig2}(b) and \ref{Fig2}(c). One can find that the primary and secondary plateaus are determined by the interband contributions of C2-V (C1-V) and C3-V, respectively. By comparing with the M--X direction, the change of transition rule gives rise to the emergence of double-plateau structure along M--$\Gamma$, as shown by the red dash-dotted line in Fig. \ref{Fig2}(a).

\begin{figure*}[htbp]
\centering
\includegraphics[width=13 cm,height=4.3 cm]{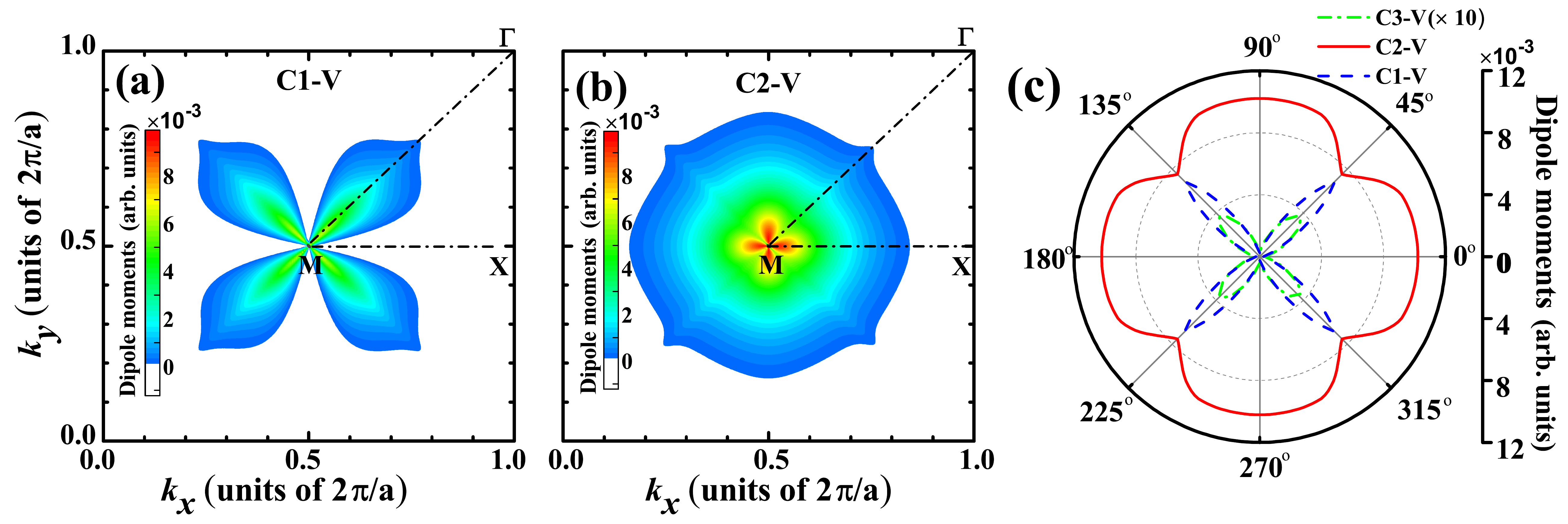}
\caption{(a) and (b) are the absolute values of the transition dipole moments between the conduction (C1 and C2) and valence (V) bands in the selected BZ based on Eq. (\ref{E4}).  (c) The polarization-dependent modulus of the transition dipoles between the conduction (C1, C2 and C3) and valence bands around the initial crystal momentum $\textbf{\emph{k}}_{0}$ .   }\label{Fig4}
\end{figure*}

In Fig. \ref{Fig3}, we pay attention to the effect of the laser polarization on the yield of high harmonics. In Fig. \ref{Fig3}(a), one can discover that the pattern of the orientation-dependent high harmonic spectra seems to be a rotating fan and holds the C$_{4}$ symmetry of the 2D system. As the features have been illustrated in the above section, the HHG spectra characterize a rapidly decreased zone and the primary plateau structure which are dominated by the mechanisms of Zener tunneling and electron-hole recombination respectively. In Fig. \ref{Fig3}(b), we show the orientation-dependent yields of the high harmonics inside the white dash-dotted circle in Fig. \ref{Fig3}(a). Their photon energies are below the energy of the minimal band gap (6.8 eV) at the initial state $\textbf{\emph{k}}_{0}$. For the below-gap harmonics, the orientation-dependent yields are decreased rapidly and maintain the C$_{4}$ symmetry, but they become more orientation-sensitive for the higher harmonics. In addition, the orientation-dependent yields for the primary and secondary plateaus exhibit a quite different symmetry compared with the below-gap harmonics in Fig. \ref{Fig3}(b), as shown in Figs. \ref{Fig3}(c) and \ref{Fig3}(d), respectively. The orientation of the highest yield for the primary plateau is along the polarization angle $\theta \approx 15^{\circ}$, as marked by the red arrow line in Fig. \ref{Fig3}(c). Whereas, the yields for the secondary plateau shown in Fig. \ref{Fig3}(d) achieve the maximal value when the polarization angle is changed to 45$^{\circ}$ (along the M--$\Gamma$ direction). The orientation-sensitive yields in the below-gap and multi-plateau zones present the obviously different dependence on angle, which implies that the HHG mechanisms among them are different and will be explained below.

In Figs. \ref{Fig4}(a) and \ref{Fig4}(b), we present the absolute values of the transition dipole moments between the valence and conduction (C1 and C2) bands in the selected BZ. These values reveal the transition rule when the 2D solids are irradiated by the linearly polarized laser field. The rule shows that the transition of C2-V is permitted for the arbitrary direction and its transition dipoles reach the maximal value along the M--X direction ($\theta = 0^{\circ}$), whereas the transitions of C1-V and C3-V (not shown) are allowed gradually when the laser polarization deviates from the above M--X direction. To assess the role of laser polarization in the Zener tunneling process, we show the details of the absolute values of the polarization-dependent transition dipole moments between all conduction and valence bands around the initial crystal momentum in the Fig. \ref{Fig4}(c).

The nonperturbative high-order harmonics are believed to be due to radiation from laser-driven motion of carriers that have tunnel ionized in the laser field \cite{Vampa1}. The rates for both direct and photon-assisted tunneling depend on the energy gap between valence and conduction band states of the same momentum and their field-free transition dipole matrix elements. Thus, the orientation-dependent yields of the below-gap harmonics shall been determined by the polarized sensitivity of the transition dipoles for C2-V around the initial momentum $\textbf{\emph{k}}_{0}$, as shown by Fig. \ref{Fig3}(b) and the red solid line in Fig. \ref{Fig4}(c), respectively. For the same regime, the orientation-dependent transition dipoles dominating the electron populations of conduction bands further lead to the modulation of HHG yields in the double-plateau zone. Just as the time-frequency analysis and quasi-classical prediction in Figs. \ref{Fig2}(b) and \ref{Fig2}(c), the harmonic bursts in the primary and secondary plateaus are contributed by the interband transitions of C2-V and C3-V, respectively. Therefore, the envelopes of the orientation-dependent yields for the primary and secondary plateaus in the Figs. \ref{Fig3}(c) and \ref{Fig3}(d) reach a good agreement with the symmetric structures for the orientation-sensitive values of the transition dipoles in Fig. \ref{Fig4}(c), as illustrated by the red solid line for C2-V and the green dash-dotted line for C3-V, respectively.

Compared with the absolute values of transition dipoles of C2-V shown by the red solid line in Fig. 4(c), however, the orientation-dependent yields of the primary plateau in Fig. \ref{Fig3}(c) characterize more fine patterns, which appears in some figures of the previously experimental and theoretical studies \cite{You_NP,ShichengJiang,Jin}, but without any discussions about them. The emergence of these fine patterns implies that the orientation-dependent transitions open the quantum-path interference between currents inside the solid induced by the laser-solid interaction.

\begin{figure}[htbp]
\centering
\includegraphics[width=8 cm,height=9 cm]{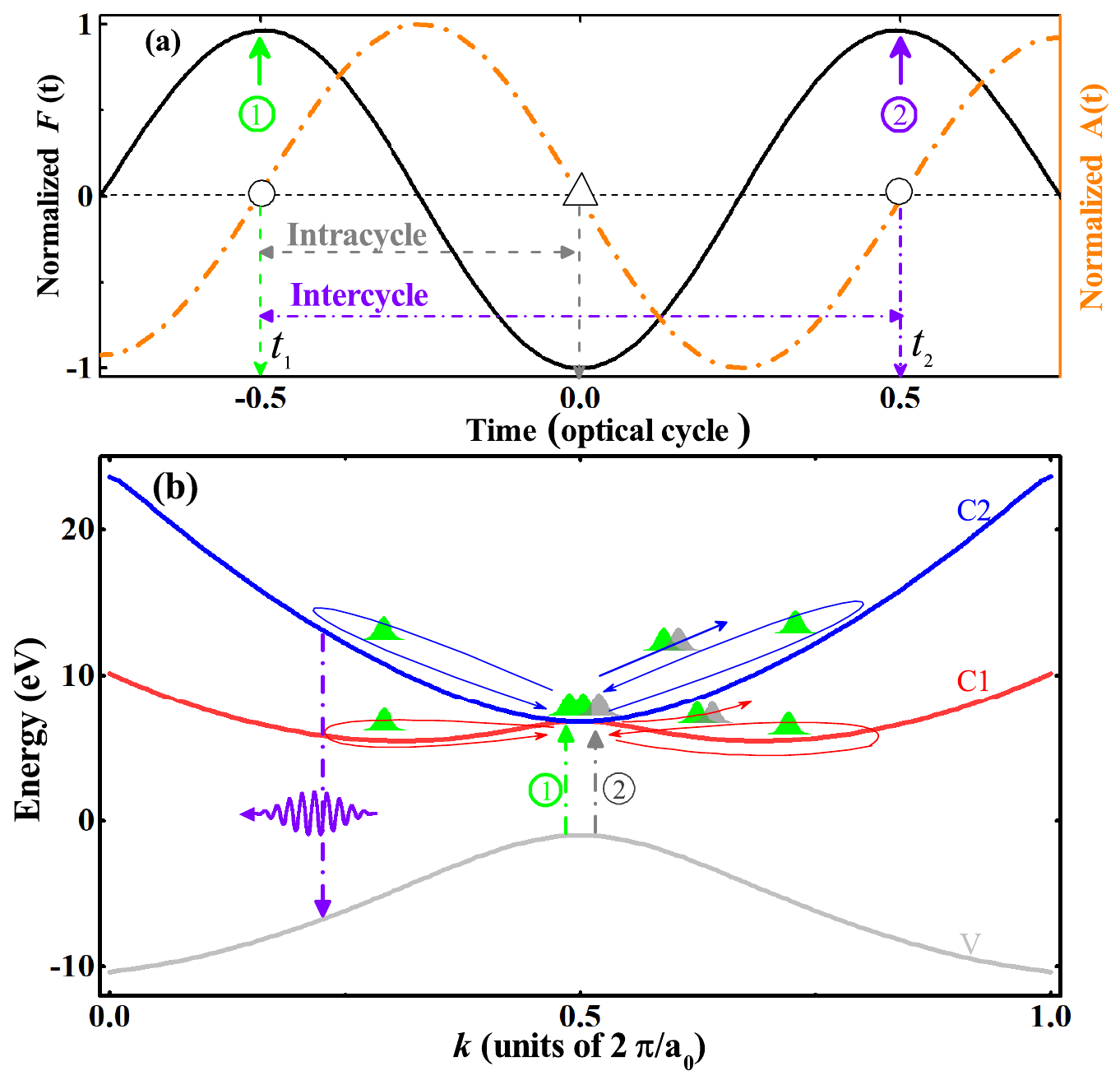}
\caption{ Schematic diagram of the intra- and intercycle interference schemes. Two electron wave packets are promoted to the conduction band by Zener transition around the two peaks which are separated by a half cycle or a full cycle. Temporal evolutions of the wave vector $\textbf{A}(t)$ is shown by the orange dash-dotted curve in (a). The motion of wave packets in the same conduction band shows the intraband current (solid arrows), while two ionization events (dashed arrows) between valence and conduction bands illustrate the Zener transitions separated by an optical cycle  in (b). The interference comes from the overlap between these groups of the wave packets. a$_{0}$ is the effective lattice constant.  }\label{Fig5}
\end{figure}

\begin{figure}[htbp]
\centering
\includegraphics[width=8.5 cm,height=11.5 cm]{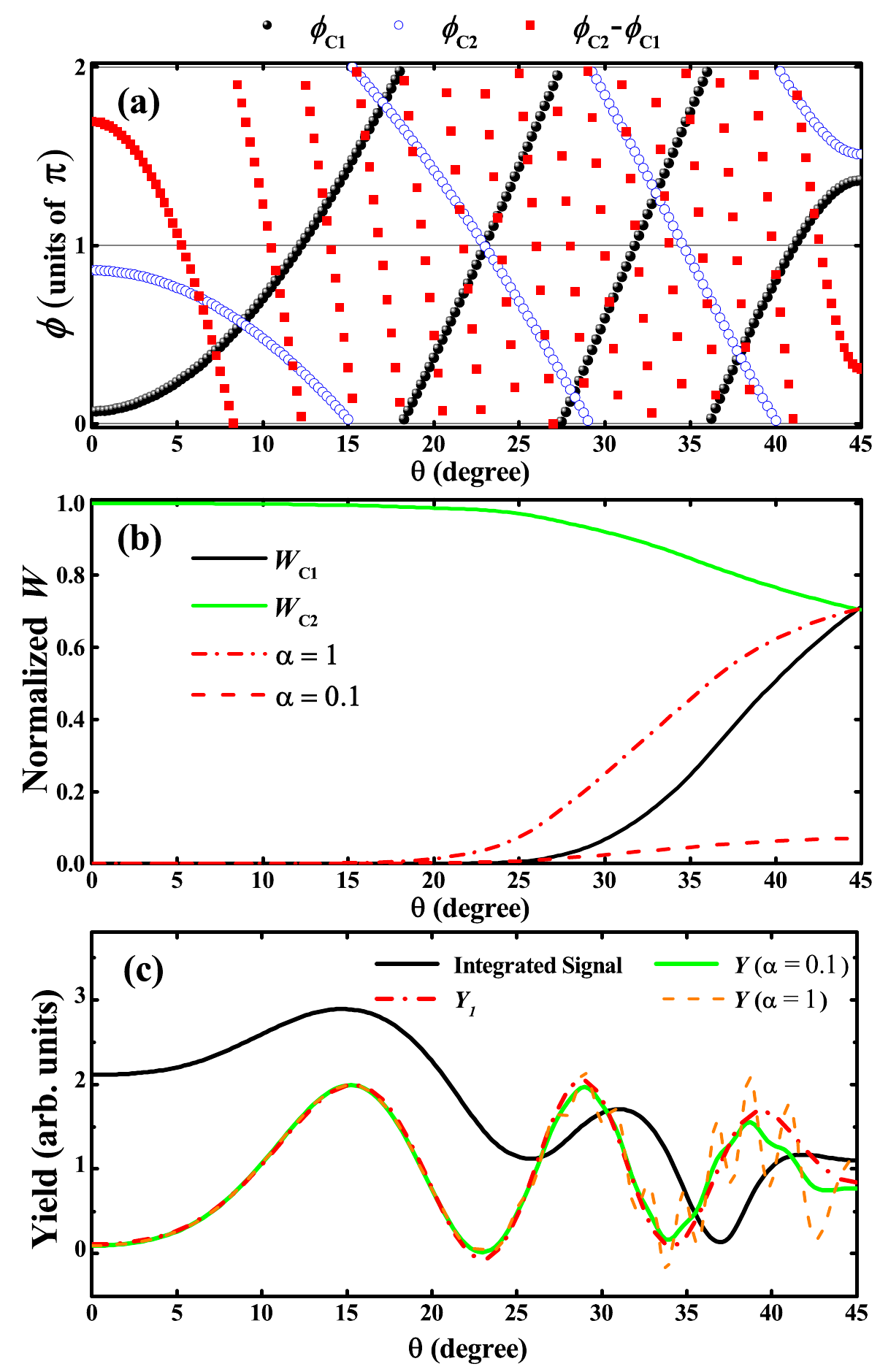}
\caption{(a) Phases of the two conduction bands and their phase difference between two transition events separated by a full laser cycle. (b) $\emph{W}_{C1}$, $\emph{W}_{C2}$ and $\alpha\sqrt{\emph{W}_{C1}\emph{W}_{C2}}$ are the normalized factors of $Y_{1}$ and $Y_{2}$ in Eq. (\ref{E7}), respectively.  (c) Comparison of the oscillation behavior of HHG yields between the model predictions (green solid line) and the integrated yields (black solid line) shown in Fig. \ref{Fig3}(c).  }\label{Fig6}
\end{figure}

In Fig. \ref{Fig5}, we propose a scheme with quantum-path interference between the laser cycles in the BZ. Firstly, let us qualitatively describe the interference process. It can be described as follows: (1) the first Zener
tunneling around the initial crystal momentum $\textbf{\emph{k}}_{0}$; (2) electrons in C1 and C2 are pushed toward $\textbf{\emph{k}}_{0} + \textbf{A}_{max}$ and then pulled back to $\textbf{\emph{k}}_{0}$ within a half cycle; (3) electrons further driven by the laser fields could reach $\textbf{\emph{k}}_{0} - \textbf{A}_{max}$ and come back to $\textbf{\emph{k}}_{0}$ again within an extra half cycle; (4) the second Zener transition event; (5) two pathways separated by an optical cycle interfere with each other according to the phase accumulated by the interband polarization between two transition events.
For a particular electron with initial crystal momentum $\textbf{\emph{k}}_{0}$, this phase is given by \cite{Shevchenko,Du5,Wolkow}
\begin{equation}\label{E6}
\phi_{m} = \frac{\pi}{2}(1-sgn[\frac{\textbf{F}(t_{2})}{\textbf{F}(t_{1})}]) +  \int_{t_{1}}^{t_{2}} \Delta E_{m}[\textbf{\emph{k}}(t)]dt,
\end{equation}
where $\Delta E_{m\in \{C1,C2\}}$ is the energy difference between the conduction and valence bands. The first and second terms on the right-hand side of this equation denote the phase differences of transition and propagation, respectively. The propagating phase is analogous to the Volkov phase in atomic physics \cite{Wolkow}. The time-dependent vector potential $\textbf{\emph{k}}(t) = \textbf{\emph{k}}_{0} + \textbf{A}(t)$. $t_{2} - t_{1} = T$, $T$ is the optical period. The transition phase difference disappears if $t_{2}$ and $t_{1}$ are separated by a full laser cycle, it is equal to zero if the laser field does not change its sign between these two moments of time. Two representative moments of time $(t_{1}, t_{2} ) \equiv (-0.5, 0.5)$ o.c. are adopted here. The phase for each conduction band and phase difference between two conduction bands are shown in Fig. \ref{Fig6}(a).

Based on the intercycle interference model, it is possible to provide an analytical calculation of the interference patterns. Combining the above polarization-sensitive transition dipoles and the phase differences between two transition events for each conduction band $m$, one can infer that the resultant yield of the harmonic signal $Y$ satisfies
\begin{equation}\label{E7}
\begin{split}
 &  Y = Y_{1} + Y_{2},  \\
 &  Y_{1} \propto 1  + \sum_{m}\emph{W}_{m} \cos\phi_{m}, \\
 &  Y_{2} \propto \sum_{m\neq m'}\alpha\sqrt{\emph{W}_{m}\emph{W}_{m'}} \cos(\phi_{m} - \phi_{m'} ), 
\end{split}
\end{equation}
where $\emph{W}_{m}$ is proportional to the electron population in conduction band $m$, i.e., tunneling rate $\gamma$, which is predicted by $\gamma \propto exp\{-\frac{2\Delta_{m}}{3|\textbf{F}||D_{m}|} \}$ and can trace back to Zener's theory on tunneling in solids \cite{Zener,Gertsvolf,Marder,Vampa_PRB}. $\Delta_{m}$ and $|D_{m}|$ are the band gap and the absolute values of transition dipoles between conduction $m$ ($\in \{C1,C2\}$) and valence bands around the initial state $\textbf{\emph{k}}_{0}$, respectively. $|D_{m}|$ is shown by the red solid and blue dash lines in Fig. \ref{Fig4}(b). In Eq. (\ref{E7}), $Y_{1}$ denotes the yields contributed by the quantum-path interference for each conduction band $m$ between two transition events separated by a full cycle. In addition, as shown by two green electron wave packets around the bottom of the two lowest conduction bands in Fig. \ref{Fig5}(b), one can find an occurrence of the overlap for these two packets which are previously promoted into the bottom by the first transition event and then propagated in their respective C1 and C2 bands independently. In Eq. (\ref{E7}), $Y_{2}$ will represent the modulated yields contributed by the overlap of these two green packets around the bottom. One assumes that two beams of light, $\sqrt{\emph{W}_{m}}e^{i\phi_{m}}$ and $\sqrt{\emph{W}_{m'}}e^{i\phi_{m'}}$, interfere with each other, in which their interference term will be delicately described by the term:  $\sqrt{\emph{W}_{m}\emph{W}_{m'}}\cos(\phi_{m}-\phi_{m'})$. However, one can further observe that the independent propagation of the two green packets in C1 and C2 bands after the overlap of the bottom occurring at the moment of the second transition event, in which the case is different from the interference between the green and gray packets overlapping continuously with each other. A scaling parameter $\alpha$ in $Y_{2}$ should been introduced to assess the magnitude of the interference between C1 and C2 bands. The contributions of $Y_{2}$ in $Y$ can be controlled by the parameter $\alpha$, as presented by the red dash ($\alpha = 0.1$) and red dash-dotted ($\alpha = 1$) lines in Fig. \ref{Fig6}(b). We make an assessment on their results shown by the green solid ($\alpha = 0.1$) and orange dash ($\alpha = 1$) curves in Fig. \ref{Fig6}(c). The parameter $\alpha = 0.1$ is adopted here and $Y_{1}$ plays the key role in the fine modulation of the orientation-dependent yields ($Y$).

Considering the symmetry shown in Fig. \ref{Fig3}(c) in mind, we just show the orientation-dependent HHG yields when the orientation angle $\theta$ changes from $0^{\circ}$ to $45^{\circ}$. In Fig. \ref{Fig6}(c), we compare the analytical results of Eq. (\ref{E7}) (green curve) with that of the integrated signal calculated by TDSE (black curve). One finds that the positions of their constructive or destructive points in the two curves agree with each other very well.A small discrepancy can be observed for a bigger angle. This is due to the fact that the detailed population distributions in a certain cycle have already been affected by the interferences caused by the previous cycles, which is not included in the above consideration \cite{Jin2}. Another possible reason is that the complicated interference events caused by the different maximum vector potentials in each half or full cycle, which are beyond the scope of this work. Finally, one may be concerned about the intracycle interference between two transition events separated by a half optical cycle, as shown in Fig. \ref{Fig5}(a). Here, we have made a check and excluded the scheme of the intracycle interference which had been discussed in Ref. \cite{Du5}.

\section{CONCLUSION}
In conclusion, we have studied the impact of orientation-dependent transition dipole on the high-order harmonic generation from solids. It leads to the emergence of the HHG multi-plateau structure through turning on the additional transition channel between the high-lying conduction and valence bands. The orientation-dependent HHG yields and its fine structure are attributed to the polarization-sensitive transition dipoles and the quantum-path interference respectively. A scheme of the intercycle interference combined with the orientation-dependent transition dipoles has been provided to make an insight into the orientational dependence and fine modulation in the HHG yields, which can map the symmetry and the two-dimensional energy band structure of the system. They can be utilized as an ultrafast tool to extract the structure of solids and a potential route for the production of coherent EUV light sources.

\section{ACKNOWLEDGMENTS}
T.-Y. Du thanks H.-H. Yang and Prof. X. B. Bian very much for helpful discussions.

\end{document}